\documentclass{osa-article}

\journal{osac}

\articletype{Research Article}

\begin{document}
\title{Observing the Effect of Polarization Mode Dispersion on Nonlinear Interference Generation in Wide-Band Optical Links}

\author{Dario Pilori,\authormark{1,2,*} Mattia Cantono,\authormark{1,3} Alessio Ferrari,\authormark{1} Andrea Carena,\authormark{1} and Vittorio Curri\authormark{1}}

\address{\authormark{1}DET, Politecnico di Torino, Corso Duca degli Abruzzi 24, Torino (TO), 10129, Italy\\
\authormark{2}Currently with INRiM, Strada delle Cacce 91, Torino (TO), 10135, Italy.\\
\authormark{3}Currently with Google LLC, 1600 Amphitheatre Parkway, Mountain View, CA 94043, U.S.A.}

\email{\authormark{*}d.pilori@inrim.it} %

\begin{abstract}
With the extension of the spectral exploitation of optical fibers beyond the
C-band, accurate modeling and simulation of nonlinear interference (NLI) generation is of the utmost performance.
Models and numerical simulation tools rely on the widely used Manakov equation (ME):
however, this approach when considering also the effect of polarization mode dispersion (PMD)
is formally valid only over a narrow optical bandwidth.
In order to analyze the range of validity of the ME and its applicability to future wide-band systems,
we present numerical simulations, showing the interplay between NLI generation and PMD over long
dispersion-uncompensated optical links, using coherent polarization division multiplexing (PDM) quadrature amplitude modulation (QAM) formats.
Using a Monte-Carlo analysis of different PMD realizations based on the coupled nonlinear Schr\"{o}dinger equations,
we show that PMD has a negligible effect on NLI generation, independently from the total system bandwidth.
Based on this, we give strong numerical evidence that the ME can be safely used to estimate NLI generation well beyond its bandwidth of validity that is limited to the PMD coherence bandwidth. 
\end{abstract}

\section{Introduction}\label{sec:intro}
To increase the capacity of coherent optical systems and 
re-configurable and transparent optical networks \cite{rizzelli_2014,Winzer:2018}
with respect the present level a viable solution is to extend the exploited optical bandwidth
beyond the currently used spectral region in the C-band \cite{Ionescu:19}.
In this transmission scenario, the optical physical layer plays a crucial role on overall performances, 
since it affects network design, management and orchestration \cite{curri2017,Auge:19}.
For this reason, predicting propagation impairments is a key enabler for performance optimization
both in the planning and in the operation phase of an optical network \cite{Filer:2018}.
It was widely shown that, with state-of-the-art transceivers based on polarization division multiplexing (PDM) and multilevel modulation formats with coherent detection, 
the main capacity-limiting effects are amplified spontaneous emission (ASE) noise introduced by optical amplifiers
and nonlinear interference (NLI) \cite{Secondini:2017}.
While ASE noise is an additive white Gaussian noise (AWGN) source, NLI, in general, is not an AWGN \cite{Dar:2013}. 
However, it was shown that, in all most common conditions, NLI can be accurately approximated as an AWGN source \cite{Poggiolini:2017}. 
This key simplifying approximation allows the development of simple and effective models to predict the power spectral density of the generated NLI \cite{poggiolini2014,carena14,dar14}.
These models are then used to derive quality of transmission (QoT) estimators, which are crucial to assess physical layer impairments of optical networks \cite{curri2017}.

All NLI models for coherent PDM optical transmission assume that fiber Kerr effect is governed by the Manakov equation (ME) \cite{marcuse1997,Agrawal:nonlinear}, 
which is an approximation of the dual-polarization coupled nonlinear Schr\"{o}dinger equation (DP-NLSE) with random birefringence.
The ME averages out the birefringence, assuming that the local orientation of its axes do not significantly vary with frequency. 
This allows the derivation of simple analytical expressions, which are exploited to derive NLI models.
However, in the presence of polarization mode dispersion (PMD), 
this equation is formally valid only over a narrow optical bandwidth, called PMD coherence bandwidth
\cite{marcuse1997,galtarossa2006,Carena:1998,McKinstrie:2004}.
Therefore, this undermines the validity of NLI models for wide-band systems.
In \cite{Serena:2011} the authors performed extensive numerical simulations on the impact of PMD
on coherent systems on dispersion-compensated (and uncompensated) systems, showing a small reduction
of NLI in the presence of PMD. However, the conditions considered in that work (number of WDM channels, pulse
shaping filter) are significantly different from modern ultra-wide-band scenarios, 
which may reduce the application of those results over such modern scenarios.
On the other hand, recent experimental demonstrations of
coherent transmission well beyond the PMD coherence bandwidth \cite{Pastorelli2012,saveedra2017,Elson:17,Cantono:JOCN:2019},
have shown a substantial agreement between models and measurements. 
This suggests that the ME, at least for transmission of coherent signals over 
long dispersion-uncompensated links, can be valid also for wide optical bandwidths,
and, consequently, we infer that PMD plays a negligible role in NLI generation.

PMD is a stochastic effect, and a thorough study of NLI generation in the presence of PMD requires Monte-Carlo analyses over a large set of
realizations. Consequently, numerical simulations are the only suitable method to perform this particular kind of analysis. 
Moreover, with numerical simulations the PMD parameter $\delta_\textup{PMD}$ can be tuned to overly large
values, even if not realistic, in order to enhance any possible PMD-induced NLI modification.

In \cite{cantono2018}, we have presented a preliminary study of NLI generation in wide optical bandwidth systems considering the presence of PMD: our results confirmed the validity of the ME over such bandwidths.
This article extends it, by providing additional results and further insights and it is organized as follows.
First, in Sec.~\ref{sec:scenario} the system scenario under analysis is described.
Then, in Sec.~\ref{sec:pmdmodel}, it is illustrated the numerical simulator based on the 
PMD \emph{coarse-step} method \cite{marcuse1997} used to simulate PMD in a split-step Fourier method (SSFM) simulation. Results are presented in Sec.~\ref{sec:res}: we analyzed propagation of a system with up to $81$ channels carrying a $32$-GBaud PDM-QAM modulations with standard $50$ GHz spacing, corresponding to a maximum optical bandwidth $B_\textup{WDM}$ of approximately $4$ THz.
We simulated the system relying either on the ME or on the coupled DP-NLSE including Monte-Carlo analyses of the PMD effect.
Finally, conclusions are delineated in Sec. \ref{sec:conc}.

\section{Simulation setup}\label{sec:simsetup}
\subsection{System scenario}\label{sec:scenario}
\begin{figure}[tbp]
\centering\includegraphics[width=0.8\linewidth]{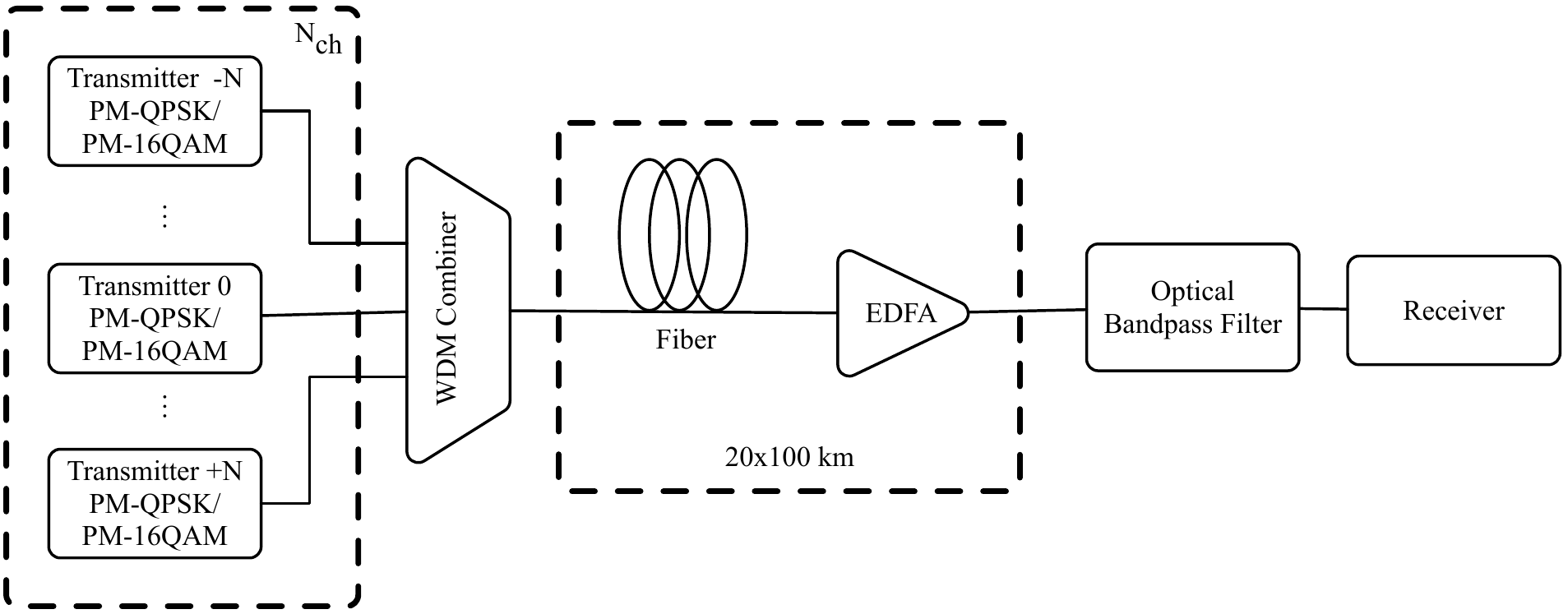}
\caption{General schematic of the simulation scenario.}
\label{fig:setup}
\end{figure}

In this work we consider a wide-band coherent transmission of PDM-QAM signals, over long dispersion-uncompensated links. 
A schematic of the analyzed link is depicted in Fig.~\ref{fig:setup}.
$N_\textup{ch}$ transmitters generate PDM-QPSK or PDM-16-QAM signals at $32$ GBaud, shaped with root-raised-cosine filters having $15\%$ roll-off.
The WDM channels, which are $50$ GHz spaced around the central frequency $f_0=193.4$ THz,
are propagated over $20\times100$-km spans of G.652 fiber (SMF) 
with typical parameters: loss of $0.2$ dB/km, dispersion of $-21.27$ ps$^2$/km and effective
area of $80~\mathrm{\mu m}^2$, giving a $1.3$ 1/(W km) nonlinear coefficient. 
After each fiber span, a flat-gain EDFA with noise figure $F = 5$ dB fully recovers span loss. 
Intra-channel stimulated Raman scattering (ISRS) is neglected, since the purpose of this work is only the analysis
of the effect of PMD on Kerr-induced NLI generation.

At the receiver, we consider as channel-under-test (CUT) the central channel of the WDM comb, and we apply a polarization and phase diversity coherent receiver. 
After photodetection and analog-to-digital conversion, signals are equalized by an LMS adaptive equalizer with $17$ taps, and the generalized 
signal-to-noise ratio (SNR) is evaluated on the  constellation.
As both channel and local oscillator lasers are assumed ideal, no carrier phase estimation (CPE) is performed.
The parameters of the equalizer were carefully chosen to have a negligible (i.e. less than $0.01$ dB) penalty, both in back-to-back
and after fiber propagation.

\subsection{Numerical simulator}\label{sec:pmdmodel}
Fiber propagation is emulated with a GPU-assisted implementation of the SSFM \cite{pilori2017,Musetti:2018}.
We analyzed the system using both the coupled DP-NLSE and the ME equations.
When applied to the DP-NLSE, the SSFM is integrated
with the coarse step method \cite{marcuse1997} to consider the random birefringence evolution that determines PMD. 
Such method approximates the continuous birefringence variations of a realistic fiber by a concatenation of fixed-length birefringent sections, each of them characterized by a random orientation of its principal states of
polarization (PSP) axis and a given differential group delay (DGD). 
PSP are a special orthogonal pairs of polarization, characterized by the fact that light launched in a PSP does not
change polarization at the output.
\cite{galtarossa2006}. 
To avoid resonance effects, we randomized the waveplate length as suggested in \cite{wai1996,marcuse1997,Eberhard2005}. 
The coarse step method is integrated in the linear step of the SSFM as described in \cite{pilori2017}.
As a reference, we also propagated the investigated signals with the PMD effect averaged out, integrating the ME.
When applying the DP-NLSE, we performed Monte-Carlo investigations of fiber propagation with PMD parameter
first set at $\delta_\textup{PMD}=0.05~\mathrm{ps}/\sqrt{\mathrm{km}}$ and then at $1~\mathrm{ps}/\sqrt{\mathrm{km}}$.
While the first value represents a typical value for modern G.652 fibers,
the latter is an overly large value that we adopted to have a polarization coherence bandwidth,
where the ME should be valid, as narrow as \cite{shtaif2000,galtarossa2006}:
\begin{equation}
\sqrt{\frac{3}{4\pi^2 \delta_\textup{PMD}^2 L_\textup{eff}}}\approx 65~\textrm{GHz}.
\label{eq:pmdcoh}
\end{equation}
This value was chosen to trigger any
possible interaction between PMD and NLI generation out of the ME validity bandwidth,
\emph{i.e.}, beyond the limits of NLI generation modeling.

\section{Results and discussion}\label{sec:res}
\begin{figure}[htbp]
\centering\includegraphics[width=0.95\linewidth]{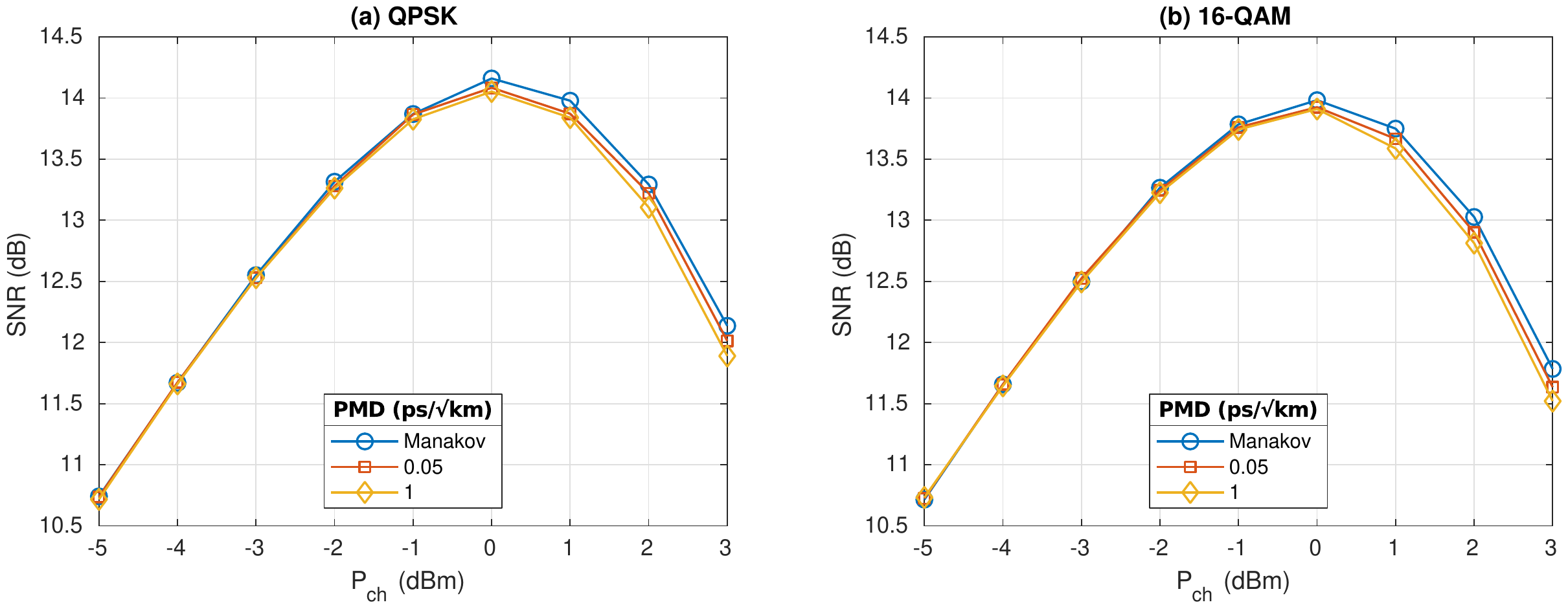}
\caption{SNR as a function of the per-channel launch power $P_\textup{ch}$
with $N_\textup{ch}=21$ WDM channels ($B_\textup{WDM}\approx 1$ THz).}
\label{fig:powcurves21}
\end{figure}
\begin{figure}[htbp]
\centering\includegraphics[width=0.95\linewidth]{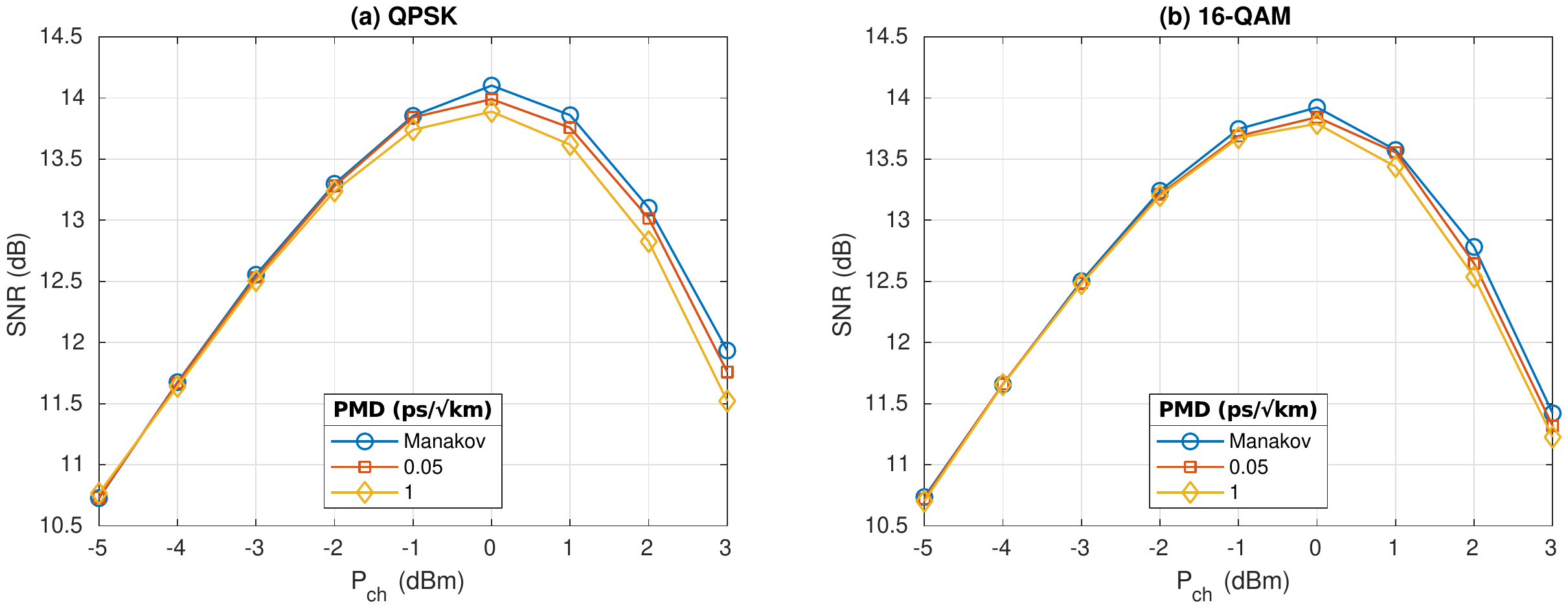}
\caption{SNR as a function of the per-channel launch power $P_\textup{ch}$
with $N_\textup{ch}=41$ WDM channels ($B_\textup{WDM}\approx 2$ THz).}
\label{fig:powcurves41}
\end{figure}

We first measured the SNR on the reference scenario ($20\times100$ km SMF) 
as a function of the per-channel launch power $P_\textup{ch}$. Results are shown in 
Fig.~\ref{fig:powcurves21} and Fig.~\ref{fig:powcurves41}  for PDM-QPSK (a) and PDM-16-QAM (b) respectively. 
One curve refer to the ME and two to the different values of $\delta_\textup{PMD}$ analyzed using the DP-NLSE.
For these first results, we measured only a single PMD realization.

In Fig.~\ref{fig:powcurves21} results are shown for the case of with $N_\textup{ch}=21$ WDM channels,
which correspond approximately to $B_\textup{WDM}=1$ THz. This value of bandwidth is larger than the PMD coherence bandwidth for
$\delta_\textup{PMD}=1~\mathrm{ps}/\sqrt{\mathrm{km}}$, whilst is smaller than PMD coherence bandwidth for
$0.05~\mathrm{ps}/\sqrt{\mathrm{km}}$ (which is $\sim 1.2$ THz).
While at low values of $P_\textup{ch}$, where performance is ASE-noise limited, all the three cases give the same SNR,
simulations with larger values of PMD give \emph{slightly} lower performances with the increase of power.
This suggests that PMD, in this case, slightly increases the amount of NLI generated.
At $P_\textup{ch}=0$ dBm, which is the optimal power value, the 
difference of SNR values obtained using ME and DP-NLSE is $0.11$ dB for PDM-QPSK and $0.07$ dB for PDM-16-QAM respectively,
for the case of $\delta_\textup{PMD}=1~\mathrm{ps}/\sqrt{\mathrm{km}}$: there is a larger increase of NLI in lower-cardinality constellations.

Then, we repeated simulations doubling the number of channels ($N_\textup{ch}=41$, $B_\textup{WDM}\approx 2$ THz):
results are shown in Fig.~\ref{fig:powcurves41}.
In this case, the optical bandwidth $B_\textup{WDM}$ is larger than the PMD coherence
bandwidth for both values of $\delta_\textup{PMD}$.
The trend is identical to the previous case when $N_\textup{ch}=21$.
The only observed difference is a small increase in the SNR gap at $P_\textup{ch}=0$ dBm and
$\delta_\textup{PMD}=1~\mathrm{ps}/\sqrt{\mathrm{km}}$, which is now $0.21$ dB for PDM-QPSK and $0.13$ dB for PDM-16-QAM.
These results strongly suggest that PMD coherence bandwidth plays a negligible role in NLI generation.

\begin{figure}[htbp]
\centering\includegraphics[width=0.5\linewidth]{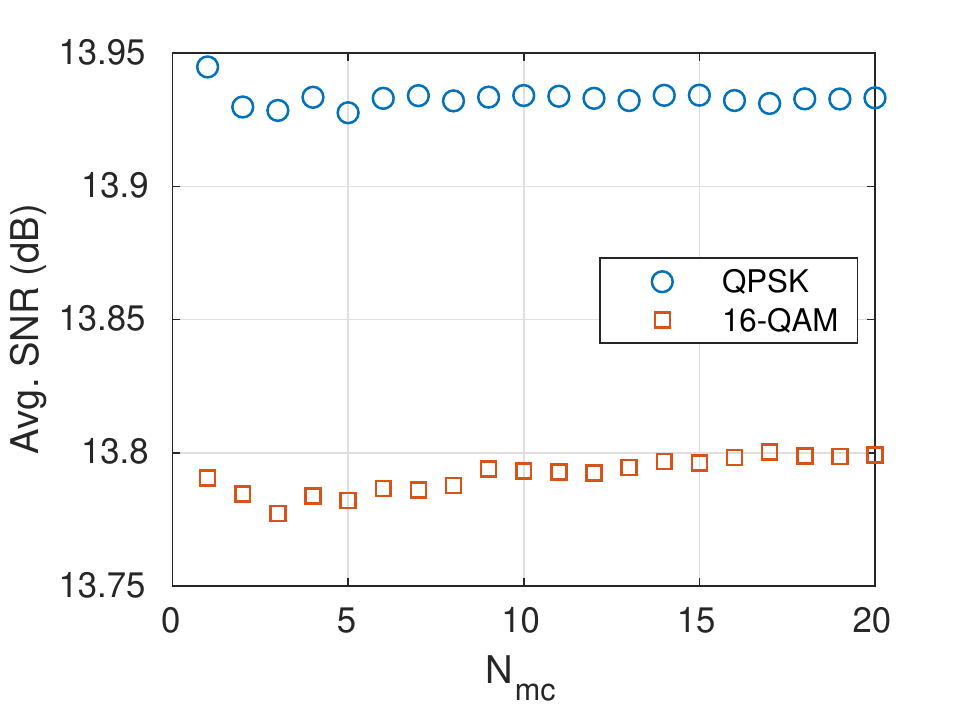}
\caption{Cumulative average of the SNR over $20$ different PMD
realizations ($N_{mc}$) at $P_\textup{ch}=0$ dBm with $N_\textup{ch}=41$ WDM channels ($B_\textup{WDM}\approx 2$ THz)
and $\delta_\textup{PMD}=1~\mathrm{ps}/\sqrt{\mathrm{km}}$.}
\label{fig:snrmontecarlo}
\end{figure}
In this latter case ($N_\textup{ch}=41$), we ran $20$ simulations and we averaged out results.
Since PMD is a stochastic effect, simulations must be verified with a Monte-Carlo analysis over different realizations:
we repeated $20$ times simulations of the $N_\textup{ch}=41$ case previously analyzed at $P_\textup{ch}=0$ dBm 
with different statistical realizations of random birefringence. 
Results are reported in Fig.~\ref{fig:snrmontecarlo}:
we show the cumulative average of the SNR at the highest PMD value 
$\delta_\textup{PMD}=1~\mathrm{ps}/\sqrt{\mathrm{km}}$.
After few realizations, the cumulative average 
converges to a stable result with an extremely low standard deviation ($<0.03$ dB). 
Consequently, we conclude that the results of Fig.~\ref{fig:powcurves41} do not depend on a specific PMD realization.

\subsection{Effect of PMD on signal statistics}
\begin{figure}[htbp]
\centering\includegraphics[width=0.9\linewidth]{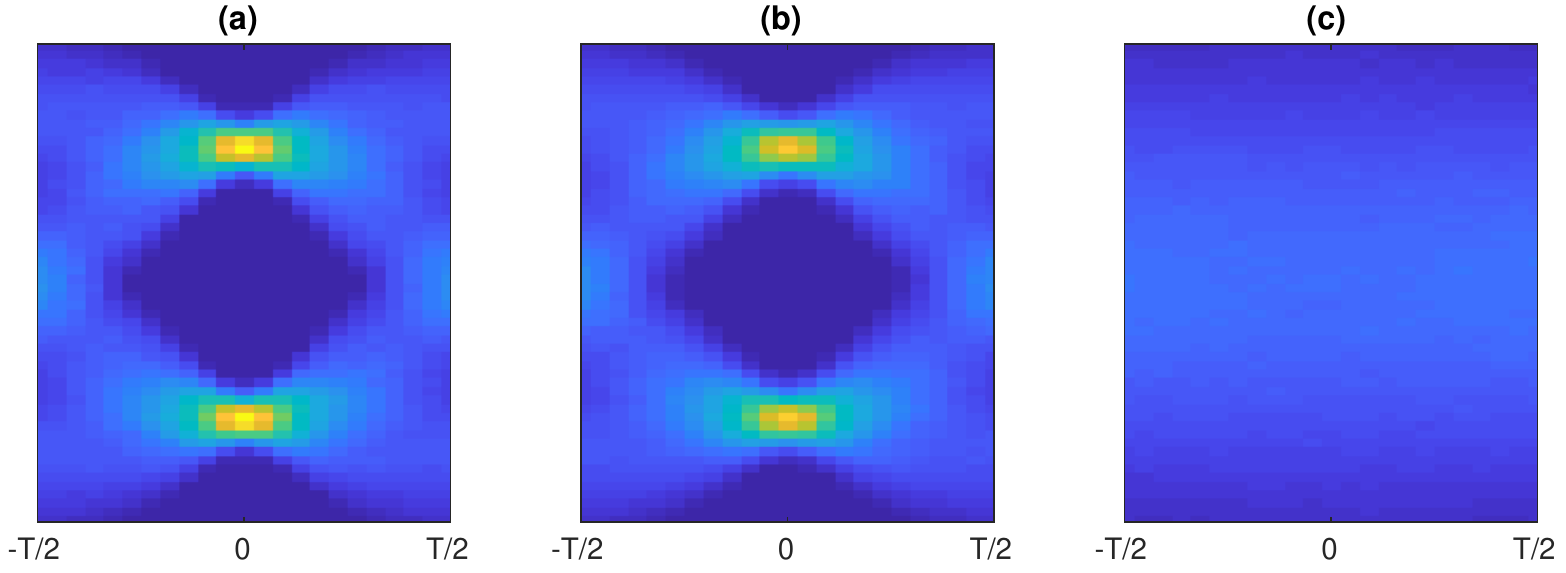}
\caption{Eye diagram of a $32$-GBaud PDM-QPSK signal at the transmitter (a) and after $23$ km of PMD-only fiber 
($\alpha=0$, $\beta_2=0$, $\gamma=0$, $\delta_\textup{PMD}=5~\mathrm{ps}/\sqrt{\mathrm{km}}$) (b) and CD-only fiber (c). $T$ is the symbol duration.}
\label{fig:qpskeyepmd}
\end{figure}
To find a possible explanation of the \emph{small} SNR difference,
we run a numerical simulation over a PMD-only fiber, \emph{i.e.} an optical fiber without attenuation, dispersion
and Kerr effect ($\alpha=0$, $\beta_2=0$, $\gamma=0$).
The simulation was run over $23$ km of fiber, which corresponds approximately to the effective length of the span.
We then measured the signal histograms before (Fig.~\ref{fig:qpskeyepmd}a) and after (Fig.~\ref{fig:qpskeyepmd}b) propagation over this optical fiber.
Since a modulated signal is a cyclostationary random process\cite{gardner2008}, with periodicity equal to the symbol duration $T$, 
one needs to create several histograms of the signal 
at different time instants within a symbol. The obtained result is very similar to an eye diagram, as shown in Fig.~\ref{fig:qpskeyepmd}.
The Figure shows $24$ different histograms of one quadrature of a $32$-GBaud PDM-QPSK signal shaped with a $15\%$ roll-off root-raised-cosine shaping filter. The number of symbols used in the simulation was $20\,000$.
For this simulation, PMD has been further increased to very large value of $5~\mathrm{ps}/\sqrt{\mathrm{km}}$ to exacerbate its effects on the signal and to allow a qualitative inspection of the histogram. 
Comparing Fig.~\ref{fig:qpskeyepmd}(a) with (b), it can be seen a slight spread of the duration of the symbol (yellow area at $T=0$).
This effect is similar (albeit much smaller) to the ``Gaussianization'' effect of chromatic dispersion (c), which makes the signal
similar to a Gaussian distribution. According to NLI generation models \cite{dar14,carena14}, a Gaussian-distributed constellation generates more NLI. Consequently, this suggests that PMD, by ``spreading'' the signal, slightly increases NLI generation, as shown in Fig.~\ref{fig:powcurves21}
and \ref{fig:powcurves41}. We remark that this effect is different from the reduction of the efficiency of 
cross phase modulation (XPM) caused by chromatic dispersion \cite{Dar:2013}, which reduces the generation of NLI.

\subsection{Propagation over low-dispersion fiber}
\begin{figure}[htbp]
\centering\includegraphics[width=0.95\linewidth]{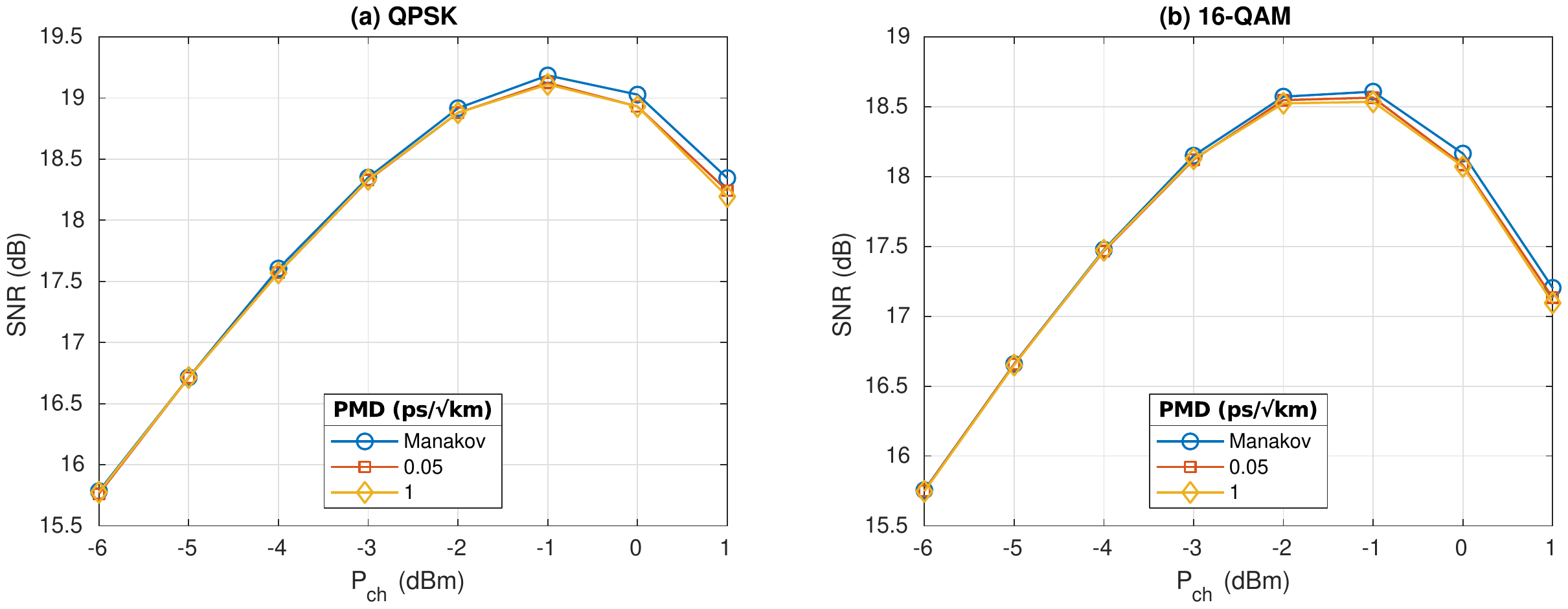}
\caption{SNR as a function of the per-channel launch power $P_\textup{ch}$
with $N_\textup{ch}=21$ WDM channels over $12\times 90$ km of NZDSF.}
\label{fig:powcurves21nzdsf}
\end{figure}
In the previous results, we have shown evidence that, in the considered scenario, PMD slightly increases
NLI generation. To give further evidence to this fact, we performed the same simulation over a different
scenario. In particular, we simulated the same setup of Fig.~\ref{fig:setup}, with $N_\textup{ch}=21$ WDM
channels, over different fiber spans. The spans were made by
standard G.655 Non-Zero Dispersion Shifted Fiber (NZDSF), with typical parameters:
attenuation $0.222$ dB/km, Kerr coefficient $1.4$ 1/(W km) and dispersion $3.8$ ps/(nm km). Span length
was reduced to $90$-km in order to have the same span attenuation as $100$-km SMF. Also the number of spans
was reduced to $12$, since we expect a stronger NLI generation, and consequently a shorter reach.
The stronger difference between this scenario, and the SMF setup, is the amount of cumulated chromatic dispersion.
In this case, chromatic dispersion is significantly lower than previous case, and it may give different
results.

Results are shown in Fig.~\ref{fig:powcurves21nzdsf}, and they are similar to SMF
results (Fig.~\ref{fig:powcurves21}). 
In fact, also in this case PMD induces a small increase of NLI, which is stronger on PDM-QPSK.
Consequently, we can conclude that the effect that is measured over SMF does not depend on that specific scenario,
but it is also present on a different scenario, with a significantly smaller amount of cumulated chromatic
dispersion.

\subsection{Extending the optical bandwidth}
\begin{figure}[htbp]
\centering\includegraphics[width=0.95\linewidth]{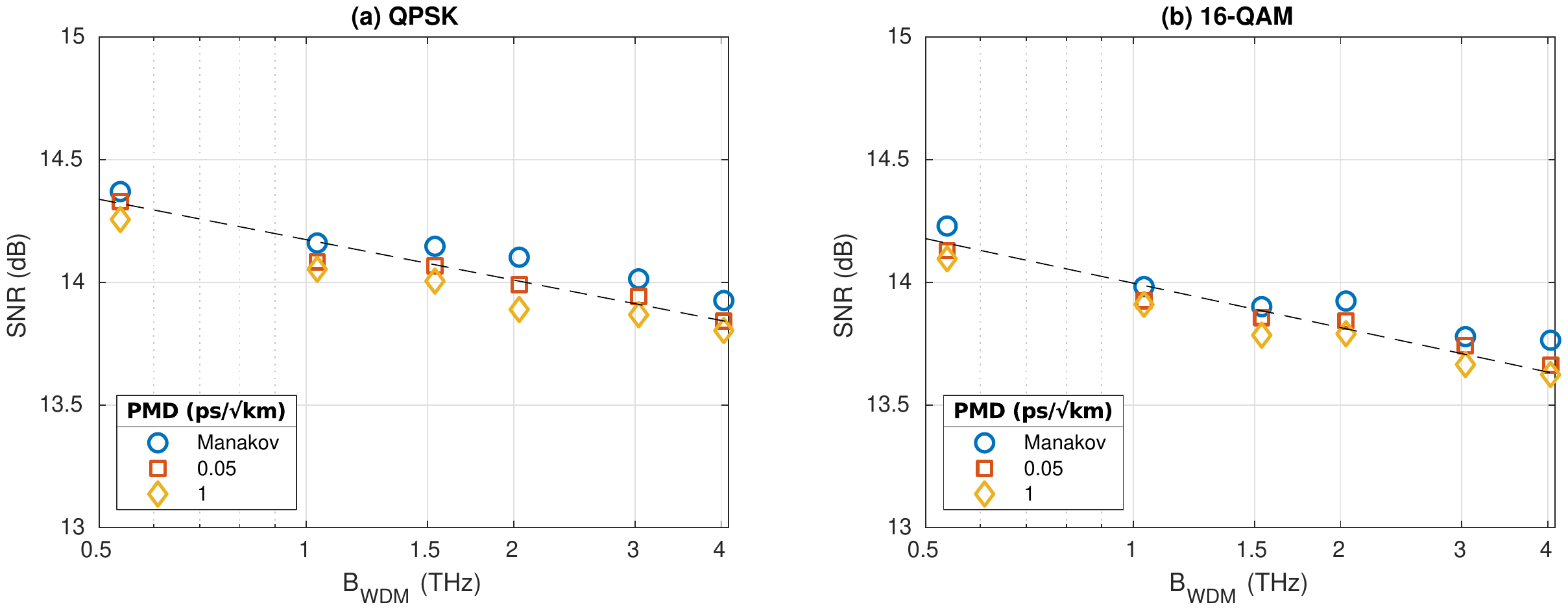}
\caption{SNR as a function of the total optical bandwidth $B_\textup{WDM}$
with fixed per-channel launch power $P_\textup{ch}=0$ dBm. Dashed line is a best-fit of an SNR decrease
proportional to the logarithm of the optical bandwidth.}
\label{fig:snrband}
\end{figure}
To investigate the bandwidth dependence of this effect,
we measured the SNR at fixed $P_\textup{ch}=0$ dBm by varying the number of WDM channels, i.e. the system optical bandwidth.
Results are shown as markers in Fig.~\ref{fig:snrband} with $N_\textup{ch}=11,21,31,41,61,81$ WDM channels.
While at low values of optical bandwidth the three results are closer,
by increasing the optical bandwidth the gap slightly increases and then it keeps approximately constant. This suggests  that this PMD-induced increase of NLI is not bandwidth-dependent.
Moreover, results in Fig.~\ref{fig:snrband} clearly indicate an increase of NLI proportional to 
$\log(B_\textup{WDM})$, which is the increase predicted by NLI models \cite{poggiolini2014,dar14,carena14} based on the ME. This can be seen by comparing in Fig.~\ref{fig:snrband} the experimental results (markers) with the black
dashed line, which is a best-fit of a linear decrease of SNR as a function of the logarithm of the optical bandwidth.
Therefore, these results suggest that, in the considered scenario, the Manakov equation is valid up to
optical bandwidths much larger than PMD coherence bandwidth. Moreover, these results are consistent with 
the experimental demonstrations presented in \cite{Pastorelli2012,saveedra2017,Elson:17}.

\section{Conclusion}\label{sec:conc}
In this paper, we presented an extensive number of simulations of wide-band transmission of PDM-QAM signals over a 
long, dispersion-uncompensated link.
We compared results obtained with the ME and with the DP-NLSE with two different values of PMD.
We found that, in this scenario, the PMD coherence bandwidth plays a negligible role in NLI generation.
Moreover, we also found that PMD \emph{slightly} increase the power of NLI. This can be due
to a change of signal statistics that enhances NLI generation.
Measuring NLI generation as a function of the optical bandwidth, we observed a logarithmic increase of NLI with bandwidth, as predicted by models based on the ME.
These results agree with recent experimental demonstrations, 
and strongly suggest that PMD plays a negligible role in NLI generation over such systems. 
However, a rigorous answer requires a thorough theoretical investigation, which is left for future research.

\section*{Funding}
The authors report no funding for this research.

\section*{Acknowledgement}
Portions of this work were presented at the OFC conference in 2018 \cite{cantono2018}.
The results presented here also appear in \cite[Sec. 2.7]{phd:cantono}.

\bibliography{bibliography}

\end{document}